\documentclass{IOS-Book-Article}

\usepackage{mathptmx}
\usepackage{soul}\setuldepth{article}
\usepackage{cite}
\usepackage{algorithmic}
\usepackage{graphicx}
\usepackage{textcomp}
\usepackage{xcolor}
\usepackage{url}
 
\def\hb{\hbox to 10.7 cm{}}

\begin{document}

\pagestyle{headings}
\def\thepage{}

\begin{frontmatter}              

\title{Second layer data governance for permissioned blockchains: the privacy management challenge}

\markboth{}{October 2020\hb}

\author[A]{\fnms{Paulo Henrique} \snm{Alves}},
\author[B]{\fnms{Isabella} \snm{Z. FRAJHOF}},
\author[A]{\fnms{Fernando A.} \snm{Correia}},
\author[A]{\fnms{Clarisse} \snm{de Souza}}
and
\author[A]{\fnms{Helio} \snm{LOPES}}

\runningauthor{P.H. Alves et al.}
\address[A]{Department of Informatics, PUC-Rio, Brazil}
\address[B]{Law Department, PUC-Rio, Brazil}

\begin{abstract}
Data privacy is a trending topic in the internet era. Given such importance, many challenges emerged in order to collect, manage, process, and publish data. In this sense, personal data have got attention, and many regulations emerged, such as GDPR in the European Union and LGPD in Brazil. This regulation model aims to protect users' data from misusage and leakage and allow users to request an explanation from companies when needed. In pandemic situations, such as the COVID-19 and Ebola outbreak, the action related to sharing health data between different organizations is/ was crucial to develop a significant movement to avoid the massive infection and decrease the number of deaths. However, the data subject, i.e., the users, should have the right to request the purpose of data use, anonymization, and data deletion. In this sense, permissioned blockchain technology emerges to empower users to get their rights providing data ownership, transparency, and security through an immutable, unified, and distributed database ruled by smart contracts. 
The governance model discussed in blockchain applications is usually regarding the first layer governance, i.e., public and permissioned models. However, this discussion is too superficial, and they do not cover compliance with the data regulations. Therefore, in order to organize the relationship between data owners and the stakeholders, i.e., companies and governmental entities, we developed a second layer data governance model for permissioned blockchains based on the Governance Analytical Framework principles applied in pandemic situations preserving the users' privacy and their duties. From the law perspective, we based our model on the UE GDPR in regard to data privacy concerns.
\end{abstract}

\end{frontmatter}
\markboth{October 2020\hb}{October 2020\hb}

\section{Introduction}
\label{sectionIntroduction}

Data privacy and data protection became one of the most critical concerns in the digital era. In order to regulate how data can be collected and used, many data protection regulations emerged to set rules to organize this environment. In Brazil, just recently, a general data protection regulation was enacted in 2018, becoming effective in September of 2021 (Law n. 13.709/2018, Lei Geral de de Prote\c{c}\~{a}o de Dados Pessoais - LGPD).  This regulation aims to provide rights and duties for both users and companies, whenever the processing of personal data is taking place. Thus, data protection norms also applies, and are extremely important in this scenario, when the processing of sensitive health data is taking place.

However, some scenarios allow the use of sensitive health data collection without the user's consent, since other legal provision can be applied; the pandemic scenario is one example\footnote{European Commission. \textit{COMMUNICATION FROM THE COMMISSION: Guidance on Apps supporting the fight against COVID 19 pandemic in relation to data protection (2020/C 124 I/01)}. Available at: \url{https://ec.europa.eu/info/sites/info/files/5_en_act_part1_v3.pdf} Accessed at: 09/20/2020.}. Data sharing and communication among health institutions, public or private, and state entities are vital to the decision making process and the definition of public policies in order to contain further disease spread \cite{Cori2017}. Moreover, the community engagement is also crucial to provide ``\textit{information delivery, consultation, collaboration in decision-making, empowering action in informal groups or formal partnerships, healthcare delivery and promotion, interaction with various stakeholders}" \cite{musesengwa2017initiating}. Previous pandemic outbreak experience like influenza, MERS-CoV, Zika\footnote{Data Sharing in Public Health Emergencies. Available at: \url{https://www.glopid-r.org/wp-content/uploads/2019/07/data-sharing-in-public-health-emergencies-yellow-fever-and-ebola.pdf} Accessed at: 10/21/2020} , and now COVID-19, showed that  data sharing between health institutions and other stakeholders worldwide are fundamental to fight against the broad contamination. The outbreak of other regional diseases in Brazil, such as Dengue, together with COVID-19, in 2021, will make the public health situation even more tricky and complex\cite{lorenz2020covid}, once they present the same clinic and laboratory characteristics \cite{chen2020epidemiological}.

From the law perspective, in Brazil, the LGPD puts forward a set of rules and obligations that regulates the use of personal data by public and private entities. Thus, in the pandemic scenario, controllers and processors, must evaluate which of the legal basis foreseen in law authorizes the collection of users' data (article 7 and 11 of the LGPD). In this sense, it must be remarked that the Brazilian data protection regulation establishes that individual consent is only one of the legal basis authorizing data processing. Furthermore, data controllers must abide to the law's principles, rights, safeguards and act in good faith. Therefore, the authorization to process data does not imply in permission to use, process, and share personal data indiscriminately. There must be a purpose for processing data, that must be legitimate, specific, explicit and previous informed to the data subject (known as the purpose principle - art. 6, I, LGPD) From a technology perspective, data privacy management is challenging. Data must be processed and kept in a safe ruled-base environment, and looks forward to a transparent and secure environment \cite{karaccam2019privacy}.

In this sense, the blockchain technology emerges as a possible solution to build a unified, distributed, trusted database. Firstly, the data immutability provided by the consensus mechanisms ensures the unified storage of historical information. Secondly, the data distribution among the worldwide network participants guarantees high data availability. Last but not least, the cryptography used in most of blockchain platforms have performed satisfying results in regards to data storage and transaction security \cite{kosba2016hawk}.

Therefore, regarding data governance, there are two main groups of blockchain platforms; the permissioned and the non-permissioned (public). The latter is broadly used for cryptocurrency applications. This model is basically full-public, i.e., anyone can read the blockchain data, and everyone can insert data since validated by the consensus protocol.
The former is regarded as the best option for second layer governance for enabling data feed and access by permissioned politics. Permissioned blockchains allow personalized data sharing; the users are able to set access rules and set which data should be public, private, or accessed under case by case authorization.

To do so, we propose a second layer of governance in permissioned blockchains solutions to fill this gap. We developed an architecture based on Hyperledger Fabric \cite{cachin2016architecture} to instantiate the proposed governance in the COVID-19 pandemic scenario. We base our model on the Governance Analytical Framework (GAF) \cite{hufty2011investigating} principles defining the Problem (such as the purpose limitation), Actors (data subject and data controller and processor), Social Norms (regulations), Process (data processor methodologies), and Nodal Points (technology used to connect stakeholders) based on the pandemic scenario.

\section{Background}
\label{sectionBackground}

\subsection{Data Regulation}
The constant and intense collection of personal data by a myriad of services and goods, and the pan-optical vigilance exercised over our behaviour when analyzing this collected data, highlights the importance of ensuring ways to protect our personal data. Due to Brazilian lack of tradition in this subject, it is important to provide society acculturation and awareness of the importance of protecting personal data in general. 

In Brazil, the LGPD puts forward a set of rules and obligations that regulates the use of personal data by public and private entities. Thus, in the pandemic scenario, controllers and processors, must evaluate which of the legal basis foreseen in law authorizes the collection of users' data (article 7 and 11 of the LGPD). In this sense, it must be remarked that the Brazilian data protection regulation establishes that individual consent is only one of the legal basis authorizing data processing. In any case, data controllers must abide to the law's principles, rights, safeguards and act in good faith. 

In this sense, we have highlighted four fundamental ideas from the LGPD, which will be applied to the scenario that we will discuss in the next sections. Firstly, the purpose limitation principle, which imposes that data processing must be legitimate, specific and explicitly informed to the data subject. Further processing is only allowed if compatible with the initial informed purpose (art. 6, I, LGPD). Secondly, the data minimization principle, meaning that it should only be used the strictly necessary data to attend to the intended and informed purpose. Thirdly, the law strongly recommends the use of anonymization or pseudonymization techniques as a governance and good practice measure to ensure data security and to protect one's privacy (arts. 12, 13, 46 and 6, VII and VIII, LGPD). Fourth, data processing must happen in a transparent manner, with the disclosure of clear, precise and easily accessible information related to the data processing activity and the controller and processor (art. 6, VI, LGPD). Furthermore, the law establishes different legal basis, beyond consent (arts. 7 and 11, LGPD), that authorizes the legitimate processing of personal data and sensitive data (which includes health data). Regardless of the legal basis used to process data, all data controllers and processors shall comply with the law's principles and other safeguards. 

In the pandemic scenario, as well as COVID-19 outbreak, the compliance with data protection norms is of the utmost importance \cite{bradford2020covid}. Such regulations do not forbid the use of personal data in such a context, but establishes the rules and legitimate uses that must be observed in pandemic scenarios. Such compliance provides that society as a whole can benefit from the uses of such data: it protects individual's privacy and data, at the same time as it allows for data utility.
In this sense, contact tracing apps \cite{ferretti2020quantifying,cho2020contact,van2020covid} are being implemented as a manner to allow public health institutions to track the infection movement and potential infected people. Some of the apps use smartphone bluetooth sensors to identify nearby devices of people that have been in contact with an infected person. In other cases, geolocation data are being being collected, although privacy concerns has been being raised. This implies the monitoring, storing and communication of the collected information, which includes sensitive information (i.e. health data). Therefore, the challenge is how to manage data privacy and comply with data protection regulations provisions.

\subsection{Blockchain Technology}
Blockchain technology emerged initially as separated concepts, block and chain, as pillar bases for creating the Bitcoin \cite{nakamoto2019bitcoin}. This technology allowed the exchange of values, assets, cryptocurrencies, and tokens between different parties without a central authority to regulate the environment. For instance, banks are not necessary anymore; people are able to do transactions peer-to-peer. Moreover, the consensus protocol delivers data immutability and empowers users to participate in the decision-making process. 

Furthermore, blockchain technology also enabled the development of smart contracts. Nick Szabo presented this concept in 1996 \cite{szabo1996smart} proposing the creation of contracts using a computer programming language with simple if-then-else logic. In 2014, Buterin \cite{buterin2014next} developed a blockchain platform that implemented the concept presented by Szabo and explored the transparency, immutability, trust, and decentralization provided by the blockchain technology. Therefore, smart contracts are immutable contracts written in programming language code that, after deployed in a blockchain environment, could be used by anyone, according to the platform data governance model. 

There are two main models of first layer blockchain governance: (i) the public (permissionless), and the (ii) private, or hybrid, (permissioned) blockchain. The former presents full data transparency, and anyone can participate in the consensus protocol \cite{crosby2016blockchain}. Moreover, anyone can do transactions since the consensus protocol validated them. There are some popular public blockchain platforms with different application domains such as: (i) Bitcoin for cryptocurrency; (ii) Ethereum for smart contracts and tokens \cite{buterin2014next}, and (iii) EOS\footnote{EOS platform. Available at: \url{https://eos.io} Accessed at: 09/20/2020.} for smart contracts and games, for instance.

The latter enables second layer data governance by the creation of policies for data access and write \cite{vukolic2017rethinking}. For this reason, the permissioned blockchains are usually applied for the corporative and governmental environments. The smart contracts created in such environment allow users to share their data under a pre-established agreement. In consequence, permissioned solutions are more compatible with data regulations worldwide. Many industries have exploring the permissioned blockchains and smart contracts, such as health \cite{ekblaw2016case}, insurance \cite{gatteschi2018blockchain}, supply-chain \cite{korpela2017digital}, oil and gas sector \cite{iceis20} and so on. It shows how the blockchain adoption has been growing in the last years in many different sectors.
\section{Data governance model}
\label{governanceModel}
In general, the data governance concept is related to big companies and how they manage a high volume of data. According to Khatri and Brown \cite{khatri2010designing}, governance is related to ``\textit{what decisions must be made to ensure effective management and use of IT and who makes the decisions.}'' Thus, data management is vital for influencing both operational and strategic decisions directly. They are crucial to the interaction and decision-making between parties that have to resolve a mutual problem. 

Even though the World Health Organization\footnote{World Health Organization. Available at: \url{https://www.who.int/} Accessed at: 09/20/2020.} (WHO) has publicly disclosed COVID-19 data, the transparency and traceability concepts have not been respected, since it is not possible to access who inserted the data, whether the data was properly anonymized, whether the patient has given informed consent, etc. Moreover, centralized platforms are subject to data unavailability for internet connection causes or hacker attack. In the pandemic outbreak, data unavailability is especially worrying and may present severe consequences to the control and management of disease spread.

Many authors discuss the challenges and opportunities in the pandemic scenario \cite{alhassan2016data,tallon2013information,hagmann2013information}, highlighting the importance of data quality. They argue that information quality depends on excellent data management, and it involves data standardization and high data availability. In summary, we selected five critical factors responsible for guaranteeing good governance in the pandemic scenario. \cite{khatri2010designing}: 
\begin{itemize}
	\item \textbf{Accountability}: related to who is the data controller, how s/he evaluated the data, and what access roles were created in regards to data management.
    \item \textbf{Standards}: related to data storage and access standards.
    \item \textbf{Partnership}: related to sharing data between organizations. 
    \item \textbf{Strategic points of control}: related to data validation and data quality assurance.
    \item \textbf{Compliance monitoring}: related to data auditing. 
\end{itemize}

Furthermore, there is a well-known approach called Governance Analytical Framework (GAF) \cite{hufty2011investigating}. The GAF is based on five principles: (i) problems, (ii) actors, (ii) social norms, (iv) processes, and (v) nodal points. This framework proposes deconstructing social problems by decomposing them on these five principles and reconstructing them by modeling the governance. This approach was used to model  COVID-19 data governance scenario. Table \ref{tableGAF} presents the association between the COVID-19 scenario and the GAF principles.

\begin{table}[htbp]
\begin{tabular}{ll}
\textbf{GAF Principle} & \textbf{COVID-19 Instantiation}                                                                                                                                                                                                                                                                                                 \\
Problem                & Health data management honoring user privacy right. \\                                                                                  
Actors                 & Actors are entities that require, or manage, the data. They are (i) data subjet (citizens) \\ & and (ii) data controller and processor (health institutions) - private and public entities.                                                                                                                                                                                        \\
Social Norms           & Social Norms are roles that define who is able to insert, update, and request data. \\ & They are: (i) data subject, whom is able to request data anytime, explanations, \\ & corrections,  and validations, and (ii) data controller and processor are able to insert data, \\ & but not exclude any  information. They must provide the data subject' answers. \\
Processes              & Processes are related to data controller and processor mechanisms applied to collect \\ & and process data.                                                                                                 \\
Nodal Points           & Nodal Points are gateways that connect data subject to data controller and data processor.                                         
\end{tabular}
\caption{COVID-19 GAF principles.} 
\label{tableGAF}
\end{table}

\begin{figure}[htbp]
\caption{User-centric model.}
\centering
\includegraphics[width=\textwidth]{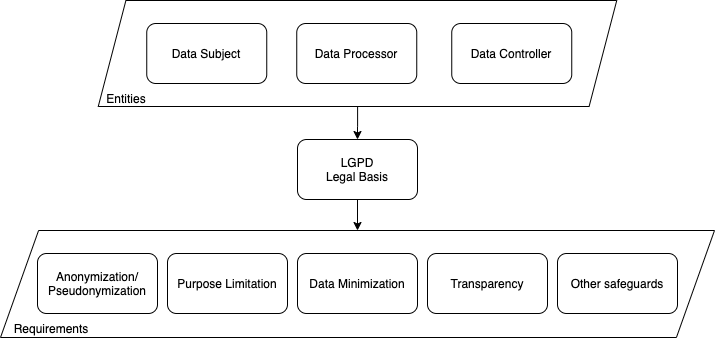}
\label{figConsent}
\end{figure}

This mapping helps people to identify the purpose of limitation accurately by verifying the Problem principle. The actors and norms involved can also be checked, so people are able to trigger, or even suite, the organization that broke any user's rights. Moreover, by checking the processes and nodal points, people can request how they were collected and processed. From the traceability perspective, the called ``\textit{contact tracing}'' apps can be modeled by the GAF principles as well. This mapping should also help health institutions to not only to elaborate explanation regarding which data will be collected, in which scenarios, in which time range, but also to guarantee data anonymization. To improve the data governance models \cite{panian2010some}, we proposed a user-centric model depicted in Figure \ref{figConsent}.

The user-centric model was based on one of the GDPR legal basis. There is six legal bases in this regulation\footnote{GDPR Consent Requirements. Available at: \url{https://gdpr.eu/gdpr-consent-requirements/} Accessed at: 09/20/2020.} and, even though the consent is one of them, it is one of the most important related to data privacy. The model highlights the actors and the consent requirements in favor of the data subject. The \textit{Collection} requirement informs that the consent must be freely given. \textit{Purpose limitation} tells that the data processor and the data controller must specify how circumstances the data will be used, and the \textit{Data Minimization} defines the minimum data to attend the expected purpose. \textit{Information} requires the action to inform who collects and processes the data in an intelligible form, using a straightforward language. \textit{Impartial Behaviour} defines that no pre-ticked box, silence, or inactivity should constitute consent. Last but not least, \textit{Revocation} means that the data subject should be able to revoke his/her consent as easy as was when it was requested. This model was proposed to enable further evolution, i.e., Entities, Requirements and the Legal Basis used can be modified to represent other regulations or focus on other concerns.

In this sense, blockchain technology was proposed as an enabler that can provide process transparency, data traceability, and empower people to be aware of the use of their data and to assert their rights. Furthermore, the permissioned blockchain also enables data sharing customization without compromising this trustful environment.

\section{Blockchain data governance}
\label{blockchainDataGovernance}

In general, blockchain governance refers to technical specifications and management. The consensus mechanisms specification, block size and block creation time are usually discussed by the developers' community in order to create different governance models. Such models are usually applied for a particular purpose, such as cryptocurrency, tokenization, digital identity, etc. However, Panian \cite{panian2010some} defined governance as ``the processes, policies, standards, organizations and technologies required to manage and ensure the availability, accessibility, quality, consistency, auditability, and security of data in an organization" and this concept may surpass the technical aspects (first layer) and drives to the society discussion, in particular, data regulations (second layer).

\begin{figure}[htbp]
\caption{GAF implementation in a permissioned blockchain architecture.}
\centering
\includegraphics[width=\textwidth]{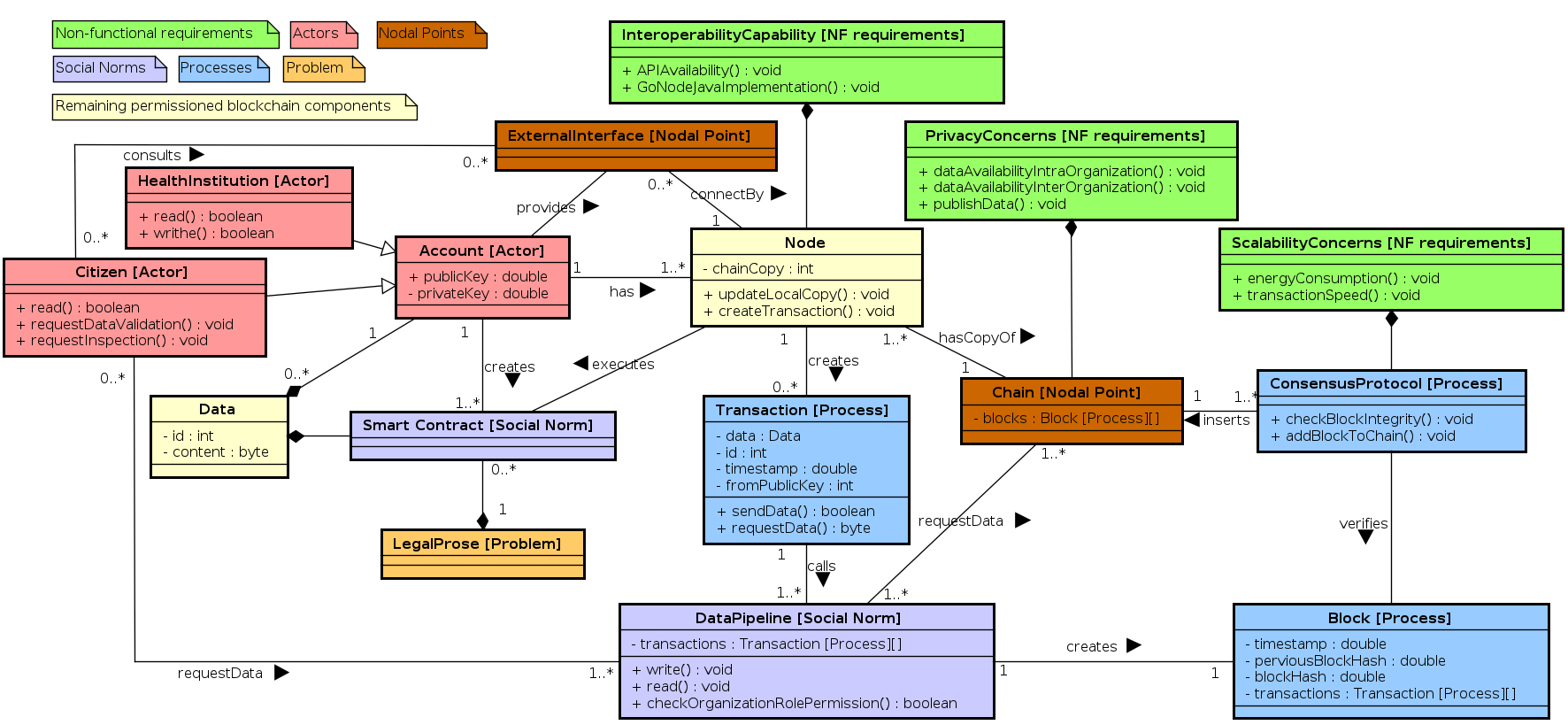}
\label{figPermissionedBlockGovernance}
\end{figure}

In order to present a complete solution to the pandemic outbreak scenario, we chose the Hyperledger Fabric permissioned blockchain to support the instantiation of the developed governance model based on the user-centric model. Permissioned blockchains fit with all the presented concepts because it allows the creation of governance rules to manage entities and data. Figure \ref{figPermissionedBlockGovernance} depicts the GAF definitions applied to this technology.

Blockchain technology provides transparency, traceability, data immutability, and availability. Moreover, permissioned blockchain adds a role layer that allows data management between selected entities. In this sense, such technology can be used to store and share pandemic data, not only as a transparent link between data subjects, data controllers, and processors, but also as a data tracker and data provider to people or any other interested entity. Through permissioned blockchain, data can be audited and be used as a data source for research purposes. Self-enforcement smart contracts enhance trust between the data subject and the data controller and processor. 

From the user-centric design perspective, the smart contracts can be used to guarantee: (i) the consent collection by requesting this information by default, (ii) purpose limitation by ensuring that such purpose will not be changed, (iii) the historical information in regards who is asking for collection and processing, (iv) the creation of standards for silence, or inactivity, data subject behavior, and (v) the timestamp\footnote{Timestamp definition:\textit{
``A record in printed or digital form that shows the time at which something happened or was done."} Available at: \url{https://dictionary.cambridge.org/dictionary/english/timestamp} Accessed at: 09/20/2020.} of revocation date.

Therefore, blockchain smart contracts play a vital role in this environment; they are responsible for roles assignment and can be used as a snapshot of activated norms in a specific moment. They are also crucial for feeding data on the blockchain. Figure \ref{figPermissionedBlockGovernance} presents the permissioned blockchain architecture based on the GAF instantiation in the pandemic scenario. We based the architecture on the Hyperledger Fabric platform concepts from the blockchain technology perspective, as did by Alketbi, Nasir, and Talib \cite{alketbi2020novel}. The presented governance architecture was developed based on GAF principles:

\textbf{Actors.} Represented by red boxes, they are: HealthInstitution (data controller and processor) and Citizen (data subject), that inherit Organization properties. The HealthInstitution is responsible for feeding health data to the blockchain and process them to be consumed by Citizen. 

\textbf{Nodal points.} Represented by brown boxes, they are ExternalInterface and Chain, which interacts directly, or indirectly, with HealthInstitutions and Citizens. They are data access points; the former can apply graphical analysis, and the latter presents the raw data. 

\textbf{Social Norms.} Purple boxes represent social norms; they are Smart Contracts and DataPipeline. Both boxes set up and verify the rights to access and write data. Moreover, smart contracts allow citizens to check the data collected regarding the data minimization principle presented by LGPD.

\textbf{Processes.} They are identified by the blue boxes, and they are represented by Transaction, Block, and ConsensusProtocol entities. These boxes manage data in order to check the primary attributes, such as timestamp and id, which are used to create the link between the blocks. 

\textbf{Problem.} This principle is represented by the LegalProse orange box, which is used to describe the scenario context in high abstraction level on the smart contract. The LegalProse also represents the purpose limitation principle foreseen in the LGPD, which means that data shall only be used to a specific purpose, previously informed and abided by the data subject, being prohibited further processing that is out of the declared scope.


In this sense, the permissioned blockchain technology structured under the GAF governance allows the second layer data governance to deal with data accountability and trustworthy data sharing by requirements in pandemic situations. Through the permissioned blockchain, people can access data from the source and verify data integrity. According to the access rules agreement on the permissioned blockchain, the data available to public consult allows researchers and governments to provide fast response in a pandemic outbreak and enables new solutions based on the user's consent. Furthermore, unified governance will enable institutions to share data following previously agreed rules. The data provenance is available for citizens, researchers, government, and health institutions, which may improve the identification of data inconsistency worldwide by information comparison.

\section{Related Work}
\label{sectionRelatedWork}

This section aims to present the works related to contact-tracing applications, privacy management models, and blockchain technology application in the health environment.

Contact tracing apps are also useful data sources for disease contamination tracking. The DP-3T initiative uses the Bluetooth signal to identify contaminated people or people that have been in touch with someone contaminated \cite{fagherazzi2020digital}. Such solution is controversial from both privacy and medical viewpoint. First, not only the infected person would be highlighted to the authorities that he/she is sick, but also people that had been in touch with him/her. Second, from the medical perspective, at least 60\% of the population would have the app installed to be effective. Therefore, to preserve the user's privacy and guarantee the necessary transparency to get people's confidence to use this kind of app, all the data should be anonymized and decentralized. In this turn, the permissioned blockchain plays a significant role in this scenario.

Panian \cite{panian2010some} argues that companies and government organizations should define standards, policies, and processes for data management. The author presents the application-centric and process-centric models for data governance.  However, the presented models do not present the concerns related to privacy and consent management required not only by the GDPR, but usually required in many other data regulations.

The authors in \cite{ekblaw2016case} proposed a blockchain-based application for electronic medical records management to deal with heavy regulations in the health sector. The blockchain smart contracts allow data sharing in this private peer-to-peer network. Even though this solution gives the immutable log, distributed information, and Accountability, they did no association with any data regulations.

As a result, the presented works showed essential concepts and applications regarding personal data collection and management. However, the junction of privacy management, governance model concepts, and the usage of blockchain technology application to provide a safe environment for data sharing has not been explored yet. Therefore, our proposal of second layer data governance for permissioned blockchain offer a promise of an environment for data sharing and privacy management.
\section{Conclusions}
\label{conclusion}

The pandemic scenario requires collaboration between citizens and public health institutions worldwide. In this sense, people must trust in the data-sharing ecosystem and their politics. The second layer data governance for permissioned blockchains aims to deliver such confidence to people feel free to give their consent. Thus, they can share their data  to contribute to define the diagnosis and evaluate methods to contain further spread.

In this paper, we proposed a new data governance for privacy management in the permissioned blockchain platforms. To do so, we used the COVID-19 outbreak scenario to apply the GAF principles to identify and define actors (data subject, controller, and processor), problems (LGPD rules), social norms, processes, and nodal points. Moreover, the LGPD rules guided our development towards compliance with data protection regulations. Such definitions were used to develop the user-centric model. This model aims to detail the concerns related to the usage of personal data maintaining compliance with data protection regulations. 

In order to apply the developed governance model, we added it as a second layer data governance in a permission blockchain technology using the Hyperledger Fabric as a platform. This technology promise to support the data subjects giving the tool to decide about sharing their data depending on the purpose limitation. Even though the permissioned blockchain is also promising to empower data subjects enabling the full control of data sharing settings, it should be deeply explored in regards to the right of rectification. As an immutable database, the rectification aspects are a big challenge as well as the right to revoke data access. These challenges are fundamental to respond to the privacy concerns and should be addressed carefully considering the available cryptography methods.

\bibliographystyle{myieeetr.bst}
\bibliography{000-main}

\end{document}